\documentclass[pra]{revtex4}

\usepackage{graphicx}
\usepackage{amssymb}
\usepackage{amsmath}
\usepackage{amsfonts}
\usepackage{epstopdf}
\usepackage{epsfig}
\usepackage{wrapfig}

\usepackage{bm}

\newcommand{\beq}{\begin{equation}}
\newcommand{\eeq}{\end{equation}}
\newcommand{\beqar}{\begin{eqnarray}}
\newcommand{\eeqar}{\end{eqnarray}}
\newcommand{\bea}{\begin{eqnarray}}
\newcommand{\eea}{\end{eqnarray}}
\newcommand{\bcen}{\begin{center}}
\newcommand{\ecen}{\end{center}}

\newcommand{\half}{\frac{1}{2}}

\begin{document}

\title{Quantum refrigerators and the III-law of thermodynamics}
\author{Amikam Levy$^{(1)}$, Robert Alicki$^{(2)(3)}$ and Ronnie Kosloff$^{(1)}$ }
\affiliation{$^{(1)}$Institute  of Chemistry  The Hebrew University, Jerusalem 91904, Israel}
\affiliation{$^{(2)}$Institute of Theoretical Physics and Astrophysics, University of
Gda\'nsk, Poland }
\affiliation{$^{(3)}$Weston Visiting Professor, Weizmann Institute of Science, Rehovot, Israel}

\begin{abstract}
The rate of temperature decrease of a cooled quantum bath is studied as its temperature is reduced to the absolute zero.
The III-law of thermodynamics is then quantified dynamically by evaluating
the characteristic exponent $\zeta$  of the cooling process $\frac{dT(t)}{dt} \sim -T^{\zeta}$ when approaching the absolute zero, $T\rightarrow 0$. A continuous model of a quantum refrigerator is employed consisting of a working medium composed either by two coupled harmonic oscillators or two coupled 2-level systems. 
The refrigerator is a nonlinear device merging three currents from three heat baths: a cold bath
to be cooled, a hot bath as an entropy sink, and a driving bath which is the source of cooling power. 
A heat driven refrigerator (absorption refrigerator) is compared to a power driven refrigerator. 
When optimized both cases lead to the same exponent $\zeta$, 
showing a lack of dependence on the form  of the working medium and the characteristics of the drivers. 
The characteristic exponent is therefore determined by the properties of the cold reservoir  and its interaction with the system. 
Two generic heat baths models are considered, a bath composed of harmonic oscillators and a bath composed from ideal Bose/Fermi gas.
The restrictions on the interaction Hamiltonian imposed by the III-law are discussed. 
In the appendix the theory of periodicaly driven open systems and its implication to thermodynamics is outlined.    
\end{abstract}

\maketitle

\section{Introduction} 
\label{sec:introduction}

Thermodynamics was initially formed as a phenomenological theory, the fundamental rules are assumed as postulates based on experimental evidence.
The well-established part of the theory concerns quasi-static macroscopic processes near thermal equilibrium. Quantum theory, on the other hand, treats the dynamical perspective of systems at atomic and smaller length scales.
Although the two disciplines reliant upon different sets of axioms, one of the first developments, Planck's law, that led to the basics of quantum theory was achieved  thanks to consistency with thermodynamics.
Einstein, following the ideas of Planck on black body radiation,  quantized the electromagnetic  field \cite{einstein05}.
\par
With the establishment of quantum theory, {\em Quantum thermodynamics} emerged as the quest to  reveal  the intimate connection between the laws of thermodynamics and their quantum origin \cite{geusic67,spohn78,alicki79,k24,k122,k156,k169,lloyd,kieu04,segal06,bushev06,erez08,mahler08,allahmahler08,segal09,he09,mahlerbook,popescu10}. 
In this tradition the present study is aimed toward the quantum study of
the third law of thermodynamics \cite{nerst06,nerst06b,nerst12,landsberg56,levy12}, in particular quantifying  the unattainability principle. 
Apart from the fundamental interest in the emergence of the third law of thermodynamics from a quantum dynamical system, 
cooling mechanical systems reveal their quantum character. 
As the temperature decreases, degrees of freedom freeze out, leaving a simplified dilute effective Hilbert space. 
Ultra-cold quantum systems contributed significantly to our understanding of basic quantum concepts. 
In addition, such systems form the basis for emerging quantum technologies. 
The necessity  to reach ultra-low temperatures requires a focus on the cooling process itself, quantum refrigeration.
\par
\begin{figure}[htbp]
\center{\includegraphics[height=6cm]{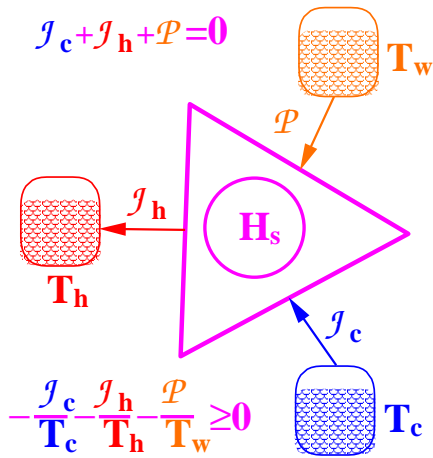}}
\caption{A quantum heat pump designated by the Hamiltonian $H_s$
coupled to a work reservoir with temperature $T_w$, a hot reservoir with temperature $T_h$ 
and a cold reservoir with temperature $T_c$. The heat and work currents are indicated. In steady state
${\cal J}_h+{\cal J}_c+{\cal P}=0$.}
\label{fig:1}
\end{figure}

The minimum requirement for constructing a continuous refrigerator is a system connected simultaneously to three reservoirs \cite{berry84}.
These baths are termed hot, cold and work reservoir  as described in Fig. \ref{fig:1}.  
This framework has to be translated to a quantum description of its components which include the 
Hamiltonian of the system $H_s$ and the implicit description of the reservoirs.
We present a careful study on the influence of different components and cooling mechanisms on the cooling process itself. 
Namely, we consider a working medium composed of two harmonic oscillators or two 2-level system (TLS). 
Two generic models of the cold heat bath are considered,
a phonons and an ideal Bose/Fermi gas heat bath. Another classification of the refrigerator is due to the character of the work reservoir. 
The first studied example is a heat driven refrigerator, an absorption refrigerator model proposed in \cite{levy12}, where $T_w \gg T_h \ge T_c$ \footnote[1]{A similar idea has been also proposed in  Phys. Rev. Lett {\bf108}, 120603 (2012) by Cleuren et.al . 
However, one can show that this model violates the III-law. The reason for this will be discused elswere.}.
In a power driven refrigerator the work reservoir represents zero entropy mechanical work, which is modeled as a periodic time dependent interaction Hamiltonian. 
\par
The models studied contain universal quantum features of such devices. 
The III-law of thermodynamics  is quantified by the characteristic exponent $\zeta$  of the change in temperature of the cold bath 
$\frac{dT_c(t)}{dt} \sim -T_c^{\zeta}$ when its temperature approaches the absolute zero, $T_c\rightarrow 0$. 
The exponent $\zeta$ is determined by a balance between the heat capacity of the cold bath and the heat current ${\cal J}_c$ into the cooling device.
When the performance of the refrigerator is optimized, the final III-law characteristics are found to be independent of the refrigerator  type.
\par
The analysis is based on a steady state operational mode of the refrigerator.
Then, the first and second laws of thermodynamics have the form:
\begin{eqnarray}
\begin{array}{rcl}
{\cal \tilde{J}}_h+{\cal \tilde{J}}_c+{\cal P}&=&0\\
-\frac{{\cal \tilde{J}}_h}{T_h}-\frac{{\cal \tilde{J}}_c}{T_c}-\frac{{\cal P}}{T_w} &\ge& 0~,
\end{array}
\label{eq:thermo}
\end{eqnarray}
where ${\cal \tilde{J}}_k $ are the stationary heat currents from each reservoir.
The first equality represents conservation of energy (first law) \cite{spohn78,alicki79}, and the second inequality 
represents non-negative entropy production in the universe $\Sigma_u \ge 0$ (second law). The fulfillment of the thermodynamical laws 
are employed to check the consistency of the quantum description. Inconsistencies can emerge either from wrong definitions of the currents 
${\cal J}_k$ or from erroneous derivations of the quantum master equation. 
In Appendix I we present a short and heuristic derivation of such a consistent Markovian master equation based on the rigorous weak coupling \cite{davies74} or low density \cite{dumcke85} limits for a constant system's Hamiltonian. 
Its generalization to periodic driving proposed in \cite{alicki06} and based on the Floquet theory is briefly discussed in Appendix II. 
Finally, in Appendix III the definition of heat currents is proposed which satisfy the second law of thermodynamics, not only for the stationary state, but also during the evolution from an arbitrary initial state of the system. It allows also to compute an averaged power in the stationary state.

\section{Quantum absorption refrigerators}
We develop and discuss in detail the model of quantum absorption refrigerator proposed in \cite{levy12}. 
We extend the results of \cite{levy12} treating by the same footing the original model with two harmonic oscillators and its two 2-level systems counterpart to stress the universality of the proposed cooling mechanism. The advantage of the absorption refrigerator is its underlying microscopic model with time-independent Hamiltonian. 
\subsection{Absorption refrigerator model }
The model consists of two harmonic oscillators or two TLS ($A$ and $B$) which are described by  two pairs of annihilation and creation operators satisfying the commutation  or anticommutation relations
\begin{equation}
a a^{\dagger} +\epsilon a^{\dagger}a =1 , a a +\epsilon a a=0 ,b b^{\dagger} +\epsilon b^{\dagger} b =1 , b b +\epsilon b b=0
\label{11}
\end{equation}
with $\epsilon = 1$ for the TLS and $\epsilon = -1$  for oscillators.
The system $A$ ($B$) is coupled to a hot(cold) bath at the temperature $T_h$ ($T_c$) and there exists a collective 
coupling of  $A+B$ to the third "work bath" at the temperature $T_w >> T_h > T_c$. 
The nonlinear coupling to the bath is essential. A linearly coupled working medium cannot operate as a refrigerator \footnote{E. Martinez and J. P. Paz, to be published.}.
The Hamiltonian of the working medium $A+B$ is given by
\begin{equation}
H = \omega_h a^+ a + \omega_c b^+ b\ , \omega_h > \omega_c
\label{12}
\end{equation}
and the interaction with the three baths (hot, cold and work) are assumed to be of the following form
\begin{equation}
H_{int} = (a + a^+)\otimes R_h +  (b + b^+ )\otimes R_c + (a b^+ + a^+ b)\otimes R_w . 
\label{12}
\end{equation}
with $R_{(\cdot)}$ being the corresponding bath operator.
\par
Applying now the derivation of the Markovian dynamics based on the weak coupling limit (see Appendix I) 
one obtains the following Markovian master equation involving three thermal generators
\begin{equation}
\frac{d\rho}{dt} = -\frac{i}{\hbar}[H , \rho ] +{\cal L}_h \rho + {\cal L}_c \rho + {\cal L}_w \rho
\label{ME_ref}
\end{equation}
where
\begin{equation}
{\cal L}_h \rho= \frac{1}{2}\gamma_h \Bigl( [a ,\rho a^{\dagger}] + e^{-\beta_h\omega_h}[a^{\dagger} ,\rho a] +h.c.\Bigr)
\label{13}
\end{equation}
\begin{equation}
{\cal L}_c \rho= \frac{1}{2}\gamma_c \Bigl( [b ,\rho b^{\dagger}] + e^{-\beta_c\omega_c}[ b^{\dagger} ,\rho b] +h.c.\Bigr)
\label{14}
\end{equation}
\begin{equation}
{\cal L}_w \rho = \frac{1}{2}\gamma_w \Bigl( [ab^{\dagger} , \rho a^{\dagger} b] + e^{-\beta_w(\omega_h-\omega_c)} [a^{\dagger} b , \rho ab^{\dagger}] +h.c.\Bigr)
\label{15}
\end{equation}
and $\beta_c > \beta_h >> \beta_w$ are inverse temperatures for the cold, hot and work bath, respectively.
\par
The values of relaxation rates $\gamma_h ,\gamma_c , \gamma_w > 0$ depend on the particular models of heat baths and their explicit form is discussed in Appendix I. Notice, that one can add also the generators describing  pure decoherence (dephasing) in the form
\begin{equation}
{\cal D}_h \rho= -\frac{1}{2}\delta_h [a^{\dagger}a ,[a^{\dagger}a , \rho]]\ ,\ {\cal D}_c \rho= -\frac{1}{2}\delta_c [b^{\dagger}b ,[b^{\dagger}b , \rho]]\ ,\ \delta_h , \delta_c >0 
\label{16}
\end{equation}
which, however, do not change the evolution of diagonal matrix elements and therefore have no influence on the cooling mechanism  at the stationary state. The generator ${\cal L}_w $ is not ergodic in the sense that it does not drive the system $A+B$ to a Gibbs state because it preserves a total number of excitations $a^{\dagger}a + b^{\dagger}b$. This
fault can be easily repaired by adding to ${\cal L}_w $ a term of the form (\ref{13}) or/and (\ref{14}) but with the temperature $T_w$. However, we assume that the processes described by (\ref{13})-(\ref{15}) dominate and additional contributions can be neglected.
\subsection{The cooling mechanism}
\label{sec:The cooling mechanism}
The  stationary  "cold current" describing heat flux from the cold bath to the working medium can be  computed using the definitions presented in Appendix III. 
The cooling of the cold bath takes place if this current is positive
\begin{equation}
\tilde{\mathcal{J}_c} =\omega_c \mathrm{Tr}\bigl[( \mathcal{L}_c\tilde{\rho})b^{\dagger}b)\bigr]> 0.
\label{17}
\end{equation}
To compute $\tilde{\mathcal{J}_c}$ we need the following equations for the mean values of the relevant observables $ {\bar n}_{h} = {\rm Tr}(\rho a^{\dagger}a)$ and ${\bar n}_{c} = {\rm Tr}(\rho b^+b)$ which can be derived using the explicit form of the generators (\ref{13}-\ref{15})
\begin{equation}
\frac{d}{dt}{\bar n}_{h} = -\gamma_h\bigl(1+\epsilon e^{-\beta_h\omega_h}\bigr){\bar n}_{h} +\gamma_h e^{-\beta_h\omega_h} +\gamma_w ({\bar n}_{c}-{\bar n}_{h}) - R
\label{18}
\end{equation}
\begin{equation}
\frac{d}{dt}{\bar n}_{c} = -\gamma_c\bigl(1+\epsilon e^{-\beta_c\omega_c}\bigr){\bar n}_{c}+\gamma_c e^{-\beta_c\omega_c} + \gamma_w ({\bar n}_{h}-{\bar n}_{c}) + R
\label{19}
\end{equation}
where $R$ is the non linear rate
\begin{equation}
R =  \gamma_w\bigl(1- e^{-\beta_w(\omega_h -\omega_c)}\bigr){\rm Tr}\bigl[\rho (b^+ b)(a a^+)\bigr]
\label{20}
\end{equation}
Eqs.(\ref{18}-\ref{20}) can be solved analytically in the high temperature limit for the work bath 
$\beta_w \to 0$ which implies $R\to 0$. Under this condition the stationary cold current reads
\begin{equation}
\tilde{\cal J}_c = \omega_c \gamma_w \frac{\bigl(e^{\beta_c\omega_c}+\epsilon\bigr)^{-1} -\bigl(e^{\beta_h\omega_h}+\epsilon\bigr)^{-1}}{1 + \gamma_w \Bigl[\gamma_h^{-1}\bigl(1+\epsilon e^{-\beta_h\omega_h}\bigr)^{-1}+ \gamma_c^{-1}\bigl(1+\epsilon e^{-\beta_c\omega_c}\bigr)^{-1}  \Bigr]}.
\label{21}
\end{equation}
The cooling condition $\tilde{\cal J}_c >0$ is equivalent to a very simple one 
\begin{equation}
\frac{\omega_c}{\omega_h} < \frac{T_c}{T_h}.
\label{22}
\end{equation}
One can similarly compute the other heat currents to obtain the coefficient of performance ($COP$)
\begin{equation}
COP = \frac{{\cal J}_c}{{\cal J}_w} = \frac{{\omega_c}}{{\omega_h - \omega_c}}.
\label{23}
\end{equation}
which becomes the Otto cycle COP \cite{mahler08,k221}.
\par
We are interested in the final stage of the cooling process when the temperature $T_c$ is close to absolute zero and hence we can assume that $\gamma_c(T_c)<< \gamma_h(T_h)$ while keeping essentially constant the value $\omega_c/T_c$. This leads to the following simplification of the formula (\ref{21})
\begin{equation}
\tilde{\cal J}_c \simeq  \omega_c  \gamma_c e^{-\omega_c/k_BT_c}.
\label{25}
\end{equation}
\section{Periodically driven refrigerator} 

Instead of driving the refrigerator by a "very hot" heat bath we apply resonant time-dependent perturbation to the system of two harmonic oscillators. 
One can repeat the derivation for 2 TLS but the final expressions for the currents are more intricate and therefore we restrict ourselves to the oscillator working medium. The time-dependent Hamiltonian reads
\begin{equation}
H(t) = \omega_h a^{\dagger} a + \omega_c b^{\dagger} b + \lambda (e^{-i\Omega t} a^{\dagger} b + e^{i\Omega t} a b^{\dagger} )\ , \Omega=\omega_h -\omega_c, \lambda >0.
\label{100}
\end{equation}
Interaction with the baths is given by
\begin{equation}
H_{int} = (a + a^{\dagger})\otimes R_h + (b+ b^{\dagger})\otimes R_c .
\label{103}
\end{equation}
The weak coupling limit Markovian master equation obtained by the methods discussed in Appendix II is of the form 
\begin{equation}
\frac{d}{dt}\rho(t) = -i[H(t),\rho(t)] + \mathcal{L}_h(t) \rho(t) +\mathcal{L}_c(t) \rho(t).
\label{ad_markov}
\end{equation}
with $ \mathcal{L}_{h(c)}(t)=  U(t,0) \mathcal{L}_{h(c)}U(t,0)^{\dagger}$ 
that can be derived directly without applying  the  full Floquet formalism. 

The main ingredients of the derivation:
\par
I) Transformation to interaction picture. The baths operators transform according to the free baths Hamiltonian, and the system operators according to the unitary propagator (under resonance conditions) 
\begin{equation}
U(t,0) = \mathcal{T}\exp\Bigl\{-i\int_0^t H(s) ds\Bigr\}= e^{-iH_0 t} e^{-iVt},
\label{101}
\end{equation}
where
\begin{equation}
H_0 = \omega_h a^{\dagger} a + \omega_c b^{\dagger} b \ ,\  V =\lambda ( a^{\dagger} b +  a b^{\dagger} )\ .
\label{102}
\end{equation}
\par
II) Fourier decomposition of the interaction part 
\begin{equation}
a(t)= U(t,0)^{\dagger}a U(t,0)= e^{iVt} \Bigl[e^{iH_0 t}a e^{-iH_0} \Bigr]e^{-iVt} = \cos(\lambda t) e^{-i\omega_h t}a - i\sin(\lambda t) e^{-i\omega_h t}b ,
\label{104}
\end{equation}
which gives the Fourier decomposition (compare with eq.(\ref{eq:Sq}))
\begin{equation}
a(t)=\frac{1}{\sqrt{2}} (e^{-i(\omega_{h}^{+}) t}d_+ + e^{-i(\omega_{h}^{-} )t}d_-)
\end{equation}
and
\begin{equation}
b(t)=\frac{1}{\sqrt{2}} (e^{-i(\omega_c^+)t}d_+ - e^{-i(\omega_c^- )t}d_-  )
\end{equation}
where $d_+=\frac{a+b}{\sqrt{2}}$, $d_-=\frac{a-b}{\sqrt{2}}$ and $\omega_{h(c)}^{\pm}=(\omega_{h(c)} \pm \lambda)$. Similarly we can calculate $a^{\dagger}(t), b^{\dagger}(t)$. 
\par
III) Performing the weak coupling approximation, the total time-independent (interaction picture) generator has the form
\begin{equation}
\mathcal{L}= \mathcal{L}_h^{(+)} + \mathcal{L}_h^{(-)}+\mathcal{L}_c^{(+)}+\mathcal{L}_c^{(-)} 
\label{106}
\end{equation}
where 
\begin{equation}
{\cal L}_{h(c)}^{(+)} \rho= \frac{1}{4}\gamma_{h(c)}^{(+)} \Bigl( [d_+ ,\rho d_+^{\dagger}] + e^{-\beta_{h(c)}\omega_{h(c)}^{+}}[d_+^{\dagger} ,\rho d_+] + h.c\Bigr) 
\label{107}
\end{equation}
and
\begin{equation}
{\cal L}_{h(c)}^{(-)} \rho= \frac{1}{4}\gamma_{h(c)}^{(-)} \Bigl( [d_- ,\rho d_-^{\dagger}] + e^{-\beta_{h(c)}\omega_{h(c)}^{-}}[d_-^{\dagger} ,\rho d_-] + h.c\Bigr) 
\label{107}
\end{equation}
with the relaxation rates $\gamma_{h(c)}^{(\pm)}=\gamma_{h(c)}(\omega_{h(c)}\pm \lambda)$ discussed explicitly in Appendices I, II. Any such generator and any sum of them possess a unique stationary state (under condition $\omega_{h(c)}\pm\lambda >0 $).
\begin{equation}
 {\tilde\rho}_{h(c)}^{(+)} = Z^{-1} exp[-\beta_{h(c)}\omega_{h(c)}^{+} d_+^{\dagger}d_+]
\end{equation}
and
\begin{equation}
 {\tilde\rho}_{h(c)}^{(-)} = Z^{-1} exp[-\beta_{h(c)}\omega_{h(c)}^{-} d_-^{\dagger}d_-]
\end{equation}
The steady (time-independent) heat currents can be computed using the definitions of the Appendix III. For example,
\begin{equation}
\tilde{\mathcal{J}}_c = - k_B T_c \Bigl[\mathrm{Tr}\bigl((\mathcal{L}_c^{(+)}{\tilde\rho})\ln {\tilde\rho}_c^{(+)}\bigr) +\mathrm{Tr}\bigl((\mathcal{L}_c^{(-)}{\tilde\rho})\ln {\tilde\rho}_c^{(-)}\bigr)\Bigr] 
\label{109}
\end{equation}
which can be calculated analytically. The result is the following
\begin{equation}
\label{Jc external field}
\tilde{\mathcal{J}}_c = \half \Biggl[\omega_c^- \frac{(e^{\beta_c \omega_c^-} -1)^{-1} -(e^{\beta_h \omega_h^-} -1)^{-1}}{[\gamma_h^{(-)}(1-e^{-\beta_h \omega_h^-})]^{-1}+[\gamma_c^{(-)}(1-e^{-\beta_c \omega_c^-})]^{-1}} +
\omega_c^+ \frac{(e^{\beta_c \omega_c^+} -1)^{-1} -(e^{\beta_h \omega_h^+} -1)^{-1}}{[\gamma_h^{(+)}(1-e^{-\beta_h \omega_h^+})]^{-1}+[\gamma_c^{(+)}(1-e^{-\beta_c \omega_c^+})]^{-1}} \Biggr]  
\end{equation}
Similarly to Sec. \ref{sec:The cooling mechanism} when the temperature $T_c$ is close to absolute zero we can assume $\gamma_c^{(-)} \ll \gamma_h^{(-)}$ and $\gamma_c^{(+)} \ll \gamma_h^{(+)}$ while keeping $\lambda/T_c < \omega_c/T_c$ as constants.
Which simplifies formula (\ref{Jc external field})
\begin{equation}
\tilde{\cal J}_c \simeq  \half \Bigl[ \omega_c^+  \gamma_c^{(+)} e^{-\omega_c^+ /k_BT_c} + \omega_c^-  \gamma_c^{(-)} e^{-\omega_c^- /k_BT_c} \Bigr].
\label{Jc external field simplified}
\end{equation}

Notice that the cold current does  not vanish when  $\lambda$   tends to zero, which obviously should be the case.
 It is due to the fact that the derivation of master equations in the weak coupling regime  involves time  averaging procedures eliminating certain oscillating terms. 
This procedure makes sense only if the corresponding Bohr frequencies are well-separated. In our case it means that  $\omega^-_c$ should be well-separated from $\omega^+_c$ which implies that  $\lambda \sim \omega_c$.  
Indeed, if both $ \omega_c$ and $\lambda $ vanish, the cold current vanishes as well.  This problem of time-scales in the weak coupling  Markovian dynamics
 has been discussed,  for constant Hamiltonians,  in \cite{alicki89} (see also \cite{davies78} for the related "dynamical symmetry breaking" phenomenon). 

\section{The dynamical III-law of thermodynamics}
\label{third law}
There exist seemingly two independent formulations of the III-law of thermodynamics, both originally stated by Nernst \cite{nerst06,nerst12}. The first is a purely static (equilibrium) one, also known as the Nernst heat theorem and can simply be phrased:

A)\emph{ The entropy of any pure substance in thermodynamic equilibrium approaches zero as the temperature approaches zero. }

The second is a dynamical one, known as the unattainability principle:

B) \emph{It is impossible by any procedure, no matter how idealized,  to reduce any assembly to absolute zero temperature in a finite number of operations} \cite{fowlerbook}.
 
Different studies investigating the relation between the two formulations, led to different answers regarding which and if at all one of these formulation imply the other. 
Although interesting, this question is out of the scope of this paper and for further considerations regarding the third law we refer the readers to \cite{landsberg56,fowlerbook,belgiorno03,belgiorno03b}. 
We shall use a more concrete version of the dynamical III-law which can be expressed as:

B') \emph {No refrigerator can cool a system to absolute zero temperature at finite time}.

This formulation enables to quantify the III-law, i.e. evaluating the characteristic exponent $\zeta$, of the cooling process $\frac{dT(t)}{dt}\sim -T^{\zeta}$ for $T\rightarrow 0$.
Namely, for $\zeta < 1$ the system is cooled to zero temperature at finite time. 
As a model of refrigerator we use the discussed above continuous refrigerators with a cold bath modeled either by a system of harmonic oscillators (bosonic bath) or the ideal gas at low density including the possible Bose-Einstein condensation effect. 
To check under what conditions the  III-law is valid, we consider a finite cold bath with the heat capacity 
$c_V(T_c)$ cooled down by the refrigerator with the optimized time-dependent parameter $\omega_c(t)$ and the additional parameter $\lambda(t)$ for the case of periodically driven refrigerator. 
The equation which describes the cooling process reads
\begin{equation}
c_V(T_c(t))\frac {d T_c(t)}{dt} = - {\cal J}_c[\omega_c(t),T_c(t)]\ , t\geq 0.
\label{24}
\end{equation}
The III-law would be violated if the solution $T_c(t)$ would reach zero at finite time $t_0$.
Now we can consider two generic models of the cold heat bath.
\subsection{Harmonic oscillator cold heat bath}
This is a generic type of a quantum bath including, for example,  electromagnetic field in a large cavity or finite but macroscopic piece of solid described in the thermodynamic limit. 
We assume the linear coupling to the bath and the standard form of the bath's Hamiltonian 
\begin{equation}
H_{int} = (b + b^{\dagger})\left(\sum_{k} (g(k)a(k) + \bar{g}(k)a^{\dagger}(k))\right)\ ,\ H_B = \sum_{k}\omega(k)a^{\dagger}(k)a(k)
\label{26}
\end{equation}
where $a(k),a^{\dagger}(k)$ are  annihilation and creation operators for a mode $k$. 
For this model the weak coupling limit procedure leads to the generator (\ref{14}) with the cold bath relaxation rate 
\begin{equation}
\gamma_c \equiv \gamma_c(\omega_c) = \pi(\sum_{k} |g(k)|^2 \delta (\omega(k) - \omega_c)[1- e^{-\omega(k)/k_BT_c}]^{-1}
\label{27}
\end{equation}
For the bosonic field in $d$-dimensional space, where $k$ is a wave vector, and with the linear  low-frequency dispersion law ($\omega(k) \sim |k|$)  we obtain  the following scaling properties at low frequencies (compare Appendix IV)
\begin{equation}
\gamma_c \sim  \omega_c^{\kappa}\omega_c^{d-1} [1- e^{-\omega_c/k_BT_c}]^{-1} 
\label{28}
\end{equation}
where $\omega_c^{\kappa}$ represents scaling of the coupling strength $|g(\omega)|^2$ and $\omega_c^{d-1}$ the density of modes scaling. It implies the following scaling
of the cold current

\begin{equation}
{\cal J}_c \sim   T_c^{d+\kappa} \Bigl[\frac{\omega_c }{T_c}\Bigr]^{d+\kappa} \frac{1}{e^{\omega_c/T_c}-1}
\label{29}
\end{equation}
Optimization of (\ref{29}) with respect to $\omega_c$ leads to the frequency tuning $\omega_c \sim T_c$ and the final current scaling
\begin{equation}
{\cal J}_c^{opt} \sim T_c^{d+\kappa}.
\label{30}
\end{equation}
Taking into account that for low temperatures the heat capacity of the bosonic systems scales like
\begin{equation}
c_{V}(T_c) \sim T_c^d
\label{31}
\end{equation}
which finally produces the following scaling of the dynamical equation (\ref{24})
\begin{equation}
\frac{d T_c(t)}{dt} \sim - (T_c)^{\kappa}.
\label{32}
\end{equation}
Notice that in a similar way the same scaling (\ref{32}) is  achieved for the periodically driven refrigerator (\ref{Jc external field simplified}), with the optimization tuning $\omega_c , \lambda \sim T_c$.\\
As a consequence, the III-law implies a rather unexpected constraint on the form of interaction with a bosonic bath 
\begin{equation}
\kappa \geq 1.
\label{33}
\end{equation}
For standard systems like electromagnetic fields or acoustic phonons with linear dispersion law $\omega(k)= v|k|$ and the form-factor $g(k)\sim |k|/\sqrt{\omega(k)}$ the parameter $\kappa =1$, as for low $\omega$, $|g(\omega)|^2 \sim |k|$. However, the condition (\ref{33}) excludes exotic 
dispersion laws $\omega(k)\sim |k|^{\alpha}$ with $\alpha < 1$ which anyway produce the infinite group velocity forbidden by the relativity theory. Moreover, the popular choice of Ohmic coupling is excluded for systems in dimension $d > 1$. The condition (\ref{33}) can be also compared with the condition 
\begin{equation}
\kappa > 2-d
\label{34}
\end{equation}
which is necessary to assure the existence of the ground state for the bosonic field interacting by means of the Hamiltonian (\ref{26})(see Appendix IV).

\subsection{Ideal Bose/Fermi gas cold heat bath}
\par
We consider now a model of a cooling process where part B of the working medium is a (infinitely) heavy particle with the internal structure approximated (at least at low temperatures) by a TLS immersed in a low density gas at temperature $T_c$. 
The Markovian dynamics of such a system was rigorously derived by Dumcke \cite{dumcke85} in the low density limit and $N$-level internal structure. The form of the corresponding LGKS generator is presented in Appendix I. For our case of TLS, we have only one Bohr frequency $\omega_c$, because elastic scattering corresponding to $\omega =0$ does not influence the cooling process. Cooling occurs due to the non-elastic scattering, giving the relaxation rate (Appendix I)
\begin{equation}
\label{gamma_ldl}
\gamma_c =2\pi n \int d^3\vec{p}\int d^3\vec{p'} \delta(E(\vec{p'}) -E(\vec{p}) - \hbar\omega_c) f_{T_c}(\vec{p}_g)|T(\vec{p'},\vec{p})|^2
\end{equation}
with $n$ the particles density, $f_{T_c}(\vec{p}_g)$ the probability distribution of the gas momentum strictly given by the Maxwell distribution, $\vec{p} $ and $\vec{p'} $ the incoming and outgoing gas particle momentum respectively. $E(\vec{p})= p^2/2m$ denotes the kinetic energy of gas particle.\\   
At low-energies (low-temperature), scattering of neutral gas at 3-d can be characterized by s-wave scattering length $a_s$, having a constant transition matrix, $|T|^2 = (\frac{4\pi a_s}{m})^2$.
For our model the integral (\ref{gamma_ldl}) is calculated
\begin{equation}
\label{gamma_ldl2}
\gamma_c= (4\pi)^4(\frac{\beta_c}{2\pi m})^{\half} a_s^2 n \omega_c {\cal K}_1(\frac{\beta_c \omega_c}{2})e^{\frac{\beta_c \omega_c}{2}} 
\end{equation}
where ${\cal K}_p(x)$ is the modified Bessel function of the second kind. 
Note that formula (\ref{gamma_ldl2}) is also valid for an harmonic oscillator instead of TLS, assuming only linear terms in the interaction and using the Born approximation for the scattering matrix.  
\par
Optimizing formula (\ref{25}) with respect to $\omega_c$ leads to $\omega_c\sim T_c$  and to scaling of the heat current
\begin{equation}
{\cal J}_c^{opt} \sim n (T_c)^{\frac{3}{2}} 
\end{equation}
When the Bose gas is above the critical temperature for the Bose-Einstein condensation the heat capacity $c_V$ and the density $n$ are constants.
Below the critical temperature the density $n$ in formula (\ref{gamma_ldl}) should be replaced by the density $n_{ex}$ of the exited states, having both $c_V, n_{ex}  \sim (T_c)^{\frac{3}{2}}$ which finally implies
\begin{equation}
\frac{dT_c(t)}{dt}\sim -(T_c)^{\frac{3}{2}}
\label{gas_scaling}
\end{equation} 
In the case of Fermi gas at low temperatures only the small fraction $n\sim T_c$ of fermions participate in the scattering process and contribute to the heat capacity, the rest is "frozen" in the "Dirac sea" below the Fermi surface.
Again, this effect modifies in the same way both sides of (\ref{24}) and therefore (\ref{gas_scaling}) is still valid.
Similarly, a possible formation of Cooper pairs below the critical temperature does not influence the scaling (\ref{gas_scaling}).
\section{Conclusions}

We have introduced and analysed  two types of continuous quantum refrigerators, an absorption refrigerator and a periodically driven refrigerator.
The latter required presenting new definitions for heat flow for periodically driven open systems, these definitions are in line with the II-law and are applicable for time independent Hamiltonian as well. 
Unlike the first and second laws, the third law of thermodynamics does not define a new state functions. 
In its first formulation (cf. section \ref{third law}) the third law provides a reference point for scaling the entropy and becomes intuitive when thinking in terms of quantum states or levels. 
The second formulation, the dynamical one $B'$ in section \ref{third law}, provides information on the characteristic exponent $\zeta$, the speed of cooling and gives an insight and restriction on properties of realistic systems.

Universal behaviour of the final scaling near the absolute zero is obtained, The III-law does not depend on the bath dimension. 
The type of refrigerator, absorption or periodically driven refrigerator, does not influence the characteristic exponent. 
Nor does a different medium, i.e. harmonic oscillator and TLS produce the same scaling. 
The characteristic exponent is governed only by the feature of the heat bath and its interaction with the system. 
For an harmonic oscillator heat bath the III-law imposes a restriction on the form of the interaction between the system and the bath, $\kappa \geq 1$. 
Allowing only physical coupling and dispersion relations, thus for phonons with a linear dispersion relation $\zeta=\kappa=1$. 
For ideal Bose/Fermi gas heat bath $\zeta=3/2$, which implies faster cooling of the phonons bath than the gas bath. 
This distinction between the two baths may occur due to particle conservation for the gas, indicating a more efficient extraction of heat by eliminating particles from the system.
The key component  of a realistic refrigerator is the heat transport mechanism between the heat bath and the working medium.
This mechanism determines the III law scaling. The working medium is a non linear device combining three currents. 
If it is optimized properly by adjusting its internal structure it does not pose a limit on cooling.         

\section*{Aknowledgements}
We want to thank Tova Feldmann, Yair Rezek, Juan Paz and Gershon Kurizki for crucial discussions.
This work is supported by the Israel Science Foundation and the Polish Ministry of Sience and Higher Education, grant number NN202208238.

\bigskip
\section{Appendix I. Thermal generators for a constant Hamiltonian}
Consider a system and a reservoir (bath), with a "bare" system Hamiltonian $H^{0}$
and the bath Hamiltonian $H_{R}$, interacting via the Hamiltonian $\lambda H_{int}=\lambda S\otimes R$.
Here, $S$ ($R$) is a Hermitian system (reservoir) operator and $\lambda $ is
the coupling strength (a generalization to more complicated $H_{int}$ is straightforward). We assume also that
\begin{equation}
[\rho_R , H_R] = 0,\ \mathrm{Tr}(\rho_R\, R) =0 .
\label{ass}
\end{equation}

\par
The reduced, system-only dynamics in the interaction picture is defined as a partial trace
\begin{equation}
\rho (t)=\Lambda (t,0)\rho \equiv \mathrm{Tr}_R \bigl(U_{\lambda}(t,0)\rho\otimes\rho_R U_{\lambda}(t,0)^{\dagger}\bigr)
\label{red_dyn}
\end{equation}
where the unitary propagator in the interaction picture is given by the ordered exponential
\begin{equation}
U_{\lambda}(t,0) = \mathcal{T}\exp\Bigl\{\frac{-i\lambda}{\hbar}\int_0^t S(s)\otimes R(s)\,ds\Bigr\}
\label{prop_int}
\end{equation}
where
\begin{equation}
S(t) = e^{(i/\hbar)Ht} S e^{-(i/\hbar)Ht} ,\  R(t)= e^{(i/\hbar)H_R t} R e^{-(i/\hbar)H_R t}.
\label{prop_int1}
\end{equation}
Notice, that $S(t)$ is defined  with respect to the renormalized, \emph{%
physical}, $H$ and not $H^{0}$ which can be expressed as
\begin{equation}
H=H^{0}+\lambda ^{2}H_{1}^{\mathrm{corr}}+\cdots .  
\label{eq:H_S}
\end{equation}
The renormalizing terms containing powers of $\lambda $ are
\emph{ Lamb-shift} corrections due to the
interaction with the bath which cancel afterwards the uncompensated term $H-H^0$ which in principle should be also present in (\ref{prop_int}).  The lowest order
(Born) approximation with respect to the coupling constant $\lambda $ yields 
$H_{1}^{\mathrm{corr}}$, while the higher order terms ($\cdots $) require
going beyond the Born approximation. 
\par
A convenient, albeit not used in the rigorous derivations, tool is  a cumulant expansion for the reduced dynamics 
\begin{equation}
\Lambda (t,0)=\exp \sum_{n=1}^{\infty }[\lambda ^{n}K^{(n)}(t)],
\end{equation}%
One finds that $K^{(1)}=0$ and the Born approximation (weak coupling) consists of terminating
the cumulant expansion at $n=2$, hence we denote $K^{(2)}\equiv K$: 
\begin{equation}
\Lambda (t,0)=\exp [\lambda ^{2}K(t)+O(\lambda ^{3})].
\end{equation}%
One obtains%
\begin{equation}
K(t)\rho =\frac{1}{\hbar^2}\int_{0}^{t}ds\int_{0}^{t}duF(s-u)S(s)\rho S(u)^{\dag }+(%
\mathrm{similar\ terms})  
\label{eq:K(t)}
\end{equation}%
where $F(s)= \mathrm{Tr}(\rho _{R}R(s)R)$.  The \emph{similar
terms} in Eq.~(\ref{eq:K(t)}) are of the form $\rho S(s)S(u)^{\dagger }$ and $S(s)S(u)^{\dagger }\rho $.
\par
The Markov approximation (in the interaction picture) means in all our cases that for long enough time one can use the following approximation
\begin{equation}
K(t)\simeq t\mathcal{L}  \label{eq:L}
\end{equation}
where $\mathcal{L}$ is a Linblad-Gorini-Kossakowski-Sudarshan(LGKS)- generator. To find its form we first decompose
$S(t)$ into its Fourier components
\begin{equation}
S(t)=\sum_{\{\omega\} } e^{i\omega t}S_{\omega }, ~~ S_{-\omega }= S_{\omega }^{\dagger}
\label{eq:S}
\end{equation}
where the set $\{\omega\}$ contains \emph{Bohr frequencies} of the Hamiltonian 
\begin{equation}
H= \sum_k \epsilon_k |k\rangle\langle k|, ~~  \omega = \epsilon_k - \epsilon_l .  
\label{Bohr}
\end{equation}
Then we can rewrite the expression (\ref{eq:K(t)}) as
\begin{equation}
K(t)\rho =\frac{1}{\hbar^2}\sum_{\omega ,\omega ^{\prime }}S_{\omega }\rho S_{\omega
^{\prime }}^{\dag }\int_{0}^{t}e^{i(\omega -\omega ^{\prime
})u}du\int_{-u}^{t-u}F(\tau )e^{i\omega \tau }d\tau +(\mathrm{similar}\text{ 
}\mathrm{terms}).  
\label{eq:K2}
\end{equation}
and use two crucial approximations:
\begin{equation}
\int_{0}^{t}e^{i(\omega -\omega ^{\prime })u}du\approx t\delta _{\omega
\omega ^{\prime }}, ~~  \int_{-u}^{t-u}F(\tau )e^{i\omega \tau }d\tau \approx {G}(\omega)=\int_{-\infty }^{\infty }F(\tau )e^{i\omega \tau }d\tau \geq 0.
\label{eq:rep1}
\end{equation}
This makes sense for $t\gg \max \{1/(\omega -\omega ^{\prime })\}$. Applying these two approximation we obtain $K(t)\rho _{S}=(t/\hbar^2)\sum_{\omega
}S_{\omega }\rho _{S}S_{\omega }^{\dag }{G}(\omega )+(\mathrm{similar}$ $%
\mathrm{terms})$, and hence it follows from Eq.~(\ref{eq:L}) that $\mathcal{L}$ is a special case of the LGKS generator derived for the first time by Davies \cite{davies74}. Returning to the Schrodinger picture one obtains the following Markovian master equation: 
\begin{eqnarray}
\frac{d\rho }{dt} &=&-\frac{i}{\hbar}[H,\rho ]+\mathcal{L}\rho ,  \notag \\
\mathcal{L}\rho  &\equiv &\frac{\lambda ^{2}}{2\hbar^2}\sum_{\{\omega \}}G(\omega
)([S_{\omega },\rho S_{\omega }^{\dagger }]+[S_{\omega }\rho ,S_{\omega
}^{\dagger }])  
\label{Dav}
\end{eqnarray}
Several remarks are in order:

\noindent (i) The absence of off-diagonal terms in Eq.~(\ref{Dav}), compared
to Eq.~(\ref{eq:K2}), is the crucial property of the Davies generator which can be 
interpreted as a coarse-graining in time of fast oscillating terms. It implies also the commutation of $\mathcal{L}$
with the Hamiltonian part $[H ,\cdot]$.

\noindent (ii) The positivity  $G(\omega )\geq 0$  follows from Bochner's theorem and is a necessary condition for
the complete positivity of the Markovian master equation.

\noindent (iii) The presented derivation showed implicitly that the notion of
\emph{bath's correlation time}, often used in the literature, is not
well-defined -- Markovian behavior involves a rather complicated cooperation
between system and bath dynamics.  In other words, contrary to what is often done in
phenomenological treatments, \emph{one cannot combine arbitrary }$H$%
\emph{'s with a given LGKS generator}. This is particularly important
in the context of thermodynamics of controlled quantum open system, where it is common to assume Markovian
dynamics and apply arbitrary control Hamiltonians. Erroneous derivations of the quantum master equation can easily lead to violation of the laws of thermodynamics.
\par
If the reservoir is a quantum system at thermal equilibrium state the additional Kubo-Martin-Schwinger (KMS) condition holds
\begin{equation}
G(-\omega) = \exp\Bigl(-\frac{\hbar\omega}{k_B T}\Bigr) G(\omega),  
\label{KMS}
\end{equation}
where $T$ is the bath's temperature. As a consequence of (\ref{KMS}) the Gibbs state 
\begin{equation}
\rho_{\beta} = Z^{-1} e^{-\beta H}, \ \beta= \frac{1}{k_B T}  
\label{gibbs atate}
\end{equation}
is a stationary solution of (\ref{Dav}). Under mild conditions (e.g : "the only system operators commuting with $H$ and $S$ are scalars") the Gibbs state is a unique stationary state and any initial state relaxes towards equilibrium ("0-th law of thermodynamics"). A convenient parametrization of the corresponding \emph{thermal generator} reads
\begin{equation} 
\mathcal{L}\rho  = \frac{1}{2}\sum_{\{\omega\geq 0\} }\gamma(\omega)\bigl\{([S_{\omega },\rho S_{\omega }^{\dagger }]+[S_{\omega }\rho ,S_{\omega}^{\dagger }]) + e^{-\hbar\beta\omega}([S_{\omega }^{\dagger},\rho S_{\omega }] +[S_{\omega }^{\dagger }\rho ,S_{\omega}])\bigr\} 
\label{Dav_therm}
\end{equation}
where finally 
\begin{equation}
\gamma(\omega)= \frac{\lambda^2}{\hbar^2} \int_{-\infty}^{+\infty} \mathrm{Tr}\bigl(\rho_R\, e^{iH_R t/\hbar}\,R\, e^{-iH_R t/\hbar}R\bigr)\, dt .  
\label{relaxation}
\end{equation}
A closer look at the expressions (\ref{Dav_therm}) (\ref{relaxation}) shows that the transition ratio from the state $|k\rangle$ to the state $|l\rangle$ is exactly the same as that computed from the Fermi Golden Rule 
\begin{equation}
W (|in\rangle \rightarrow |fin\rangle) = \frac{2\pi}{\hbar}|\langle in|V|fin\rangle |^2 \delta(E_{fin}-E_{in}) .  
\label{fermi}
\end{equation}
Namely, one should take as a perturbation $V = \lambda S\otimes R$, an initial state $|in\rangle = |k\rangle\otimes |E\rangle$,
a final state $|fin\rangle = |l\rangle\otimes |E'\rangle$ ($|E\rangle$- reservoir's energy eigenstate), and integrate over initial reservoir's states with the equilibrium distribution and over all final reservoir's states.
\par
The above interpretation allows to justify the extension of the construction of thermal generator to the case of heat bath consisting of non-interacting particles at low density $n$ and thermal equilibrium (see \cite{dumcke85} for rigorous derivation). In this case a fundamental relaxation process is a scattering of a single bath particle with the system described by the scattering matrix $T$. The scattering matrix can be decomposed as $T =\sum_{\{\omega\}} S_{\omega}\otimes R_{\omega}$ where now $R_{\omega}$ are single particle operators. Then the structure of the corresponding master equation is again given by (\ref{Dav_therm}) with
\begin{equation}
\label{gamma ldl}
\gamma(\omega) =2\pi n \int d^3\vec{p}\int d^3\vec{p'} \delta(E(\vec{p'}) -E(\vec{p}) -\hbar\omega) M(\vec{p})|T_{\omega}(\vec{p'},\vec{p})|^2
\end{equation}
resembling a properly averaged expression (\ref{fermi}). Here the initial (final) state has a structure $|k\rangle\otimes|\vec{p}\rangle$ ( $|l\rangle\otimes|\vec{p'}\rangle $), $M(\vec{p})$ is the equilibrium (Maxwell) initial distribution of particle momenta, with $|\vec{p}\rangle$ being particle momentum eigenvector, $ E(\vec{p})$ kinetic energy of a particle. The perturbation $V$ in (\ref{fermi}) is replaced by the scattering matrix $T$ (equal to $V$ for Born approximation) and finally
\begin{equation}
\label{t-matrix }
T_{\omega}(\vec{p'},\vec{p}) = \langle\vec{p'}|R_{\omega}|\vec{p}\rangle .
\end{equation}
\section{Appendix II. Thermal generators for periodic driving}
In order to construct models of quantum heat engines or powered refrigerators we have to extend the presented derivations of Markovian master equation to the case of periodically driven systems. Fortunately, we can essentially repeat the previous derivation with the following amendments:

1) The system (physical, renormalized) Hamiltonian is now periodic
\begin{equation}
\label{Ham_per}
H(t)= H(t + \tau),\ U(t,0) \equiv \mathcal{T}\exp\bigl\{-\frac{i}{\hbar}\int_0^t H(s)\,ds\bigr\},
\end{equation}
and the role of constant Hamiltonian is played by $H$ defined as
\begin{equation}
H= \sum_k \epsilon_k |k\rangle\langle k| ,\ U(\tau,0)= e^{-iHt/\hbar} .
\label{eq:Sq1}
\end{equation}

2) The Fourier decomposition (\ref{eq:S}) is replaced by the following one 
\begin{equation}
U(t,0)^{\dagger}\,S\, U(t,0)=\sum_{q\in \mathbf{Z}}\sum_{\{\omega\}} e^{i(\omega+ q\Omega) t}S_{\omega q},  
\label{eq:Sq}
\end{equation}
where $\Omega = 2\pi/\tau$  and  $\{\omega\}= \{\epsilon_k - \epsilon_l\}$.
The decomposition of above follows from the Floquet theory, however for our model we can obtain it directly using the manifest expressions for the propagator $U(t,0)$.

3) The generator in the interaction picture has form:
\begin{equation} 
\mathcal{L}  = \sum_{q\in \mathbf{Z}}\sum_{\{\omega\}}=\mathcal{L}_{\omega q}
\label{Dav_per}
\end{equation}
where
\begin{equation} 
\mathcal{L}_{\omega q}\rho  = \frac{1}{2}\gamma(\omega+q\Omega)\bigl\{([S_{\omega q},\rho S_{\omega q}^{\dagger }]+[S_{\omega q}\rho ,S_{\omega q}^{\dagger }]) + e^{-\hbar\beta(\omega+q\Omega)}([S_{\omega q}^{\dagger},\rho S_{\omega q}] +[S_{\omega q}^{\dagger }\rho ,S_{\omega q}])\bigr\} .
\label{Dav_per1}
\end{equation}
Returning to the Schrodinger picture we obtain the following master equation:
\begin{equation}
{\frac{d\rho(t) }{dt}}= -\frac{i}{\hbar}[H(t), \rho(t)]+\mathcal{L}(t)\rho(t), ~~ t\geq 0.
  \label{ME_per2}
\end{equation}
where 
\begin{equation}
\mathcal{L}(t)= \mathcal{L}(t+\tau ) =\mathcal{U}(t,0)\mathcal{L}\mathcal{U}(t,0)^{\dagger }, ~~ \mathcal{U}(t,0)\cdot
= U(t,0)\cdot U(t,0)^{\dagger}.
\label{gen_per} 
\end{equation}
In particular one can represent the solution of (\ref{ME_per2}) in the form
\begin{equation}
\rho(t) = \mathcal{U}(t,0)e^{\mathcal{L}t}\rho(0), ~~ t\geq 0.  
\label{sol}
\end{equation}
Any state, satisfying $\mathcal{L}\tilde{\rho}= 0$, defines a periodic steady state (limit cycle)
\begin{equation}
\tilde{\rho}(t) = \mathcal{U}(t,0)\tilde{\rho} = \tilde{\rho}(t+\tau), ~~ t\geq 0.  
\label{persol}
\end{equation}
Finally one should notice that in the case of multiple couplings and multiple heat baths the generator $\mathcal{L}$ can be always represented as an appropriate sum of the terms like (\ref{Dav_therm}).

\section{Appendix III. Heat flows and power for periodically driven open systems}
We consider a periodically driven system coupled to several heats with the additional index $j$ labeling them. Then the generator in the interaction picture has form
\begin{equation}
\mathcal{L}=\sum_{j=1}^M \sum_{q\in \mathbf{Z}}\sum_{\{\omega\geq 0\}}\mathcal{L}^j_{\omega q},
\label{decomposition}
\end{equation}
where any single $\mathcal{L}^j_{\omega q}$ has a structure of (\ref{Dav_per1}) with the appropriate $\gamma_j(\omega)$.
Notice, that a single component $\mathcal{L}^j_{\omega q} $ is also a LGKS generator and possesses a Gibbs-like stationary state written in terms of the averaged Hamiltonian $H$
\begin{equation}
\tilde{\rho}^j_{\omega q} = Z^{-1} \exp\Bigl\{-\frac{\omega + q\Omega}{\omega}\frac{H}{k_B T_j}\Bigr\}, 
\label{gibbs_loc}
\end{equation}
The corresponding time-dependent objects satisfy
\begin{equation}
\mathcal{L}^j_{q\omega}(t) \tilde{\rho}^j_{q\omega}(t) = 0, \ \mathcal{L}^j_{q\omega}(t) =\mathcal{U}(t,0)\mathcal{L}^j_{q\omega}\mathcal{U}(t,0)^{\dagger },\   \tilde{\rho}^j_{q\omega}(t) = \mathcal{U}(t,0)\tilde{\rho}^j_{q\omega} = \tilde{\rho}^j_{q\omega}(t+\tau) \  .
\label{inv.state}
\end{equation}
Using the decomposition (\ref{decomposition}), one can define a \emph{local heat current} which corresponds to the exchange of energy $\omega + q\Omega$ with the $j$-th heat bath for any initial state 
\begin{equation}
{\mathcal{J}^j_{q\omega}}(t) = \frac{\omega + q\Omega}{\omega}\mathrm{Tr}\bigl[( \mathcal{L}^j_{q\omega}(t)\rho(t))\tilde{H}(t)\bigr],\ \tilde{H}(t)=\mathcal{U}(t,0)H ,
\label{curr_loc}
\end{equation}
or in the equivalent form
\begin{equation}
{\mathcal{J}^j_{q\omega}}(t) = -k_B T_j\mathrm{Tr}\bigl[( \mathcal{L}^j_{q\omega}(t)\rho(t))\ln \tilde{\rho}^j_{q\omega}(t)\bigr].
\label{curr_loc1}
\end{equation}
\par
The heat current associated with the $j$-th bath is a sum of the corresponding local ones
\begin{equation}
{\mathcal{J}^j}(t) = -k_B T_j\sum_{q\in \mathbf{Z}}\sum_{\{\omega\geq 0\}}\mathrm{Tr}\bigl[( \mathcal{L}^j_{q\omega}(t)\rho(t))\ln \tilde{\rho}^j_{q\omega}(t))\bigr].
\label{curr_per1}
\end{equation}
In order to prove the II-law we use Spohn's inequality  \cite{spohn78}
\begin{equation}
\mathrm{Tr}\bigl([\mathcal{L}\rho ][\ln \rho - \ln \tilde{\rho}]\bigr) \leq 0
\label{spohn}
\end{equation}
valid for any LGKS generator $\mathcal{L}$ with a stationary state $\tilde{\rho}$.
\par
Computing now the time derivative of the entropy  $S(t)= -k_B \mathrm{Tr}\rho(t)\ln\rho(t)$ and applying (\ref{spohn}) one obtains the II-law in the form
\begin{equation}
\frac{d}{dt}S(t) - \sum_{j=1}^M \frac{\mathcal{J}^j(t)}{T_j} \geq 0
\label{IIlaw_periodic}
\end{equation}
where $S(t) = -k_B \mathrm{Tr}\bigl(\rho(t)\ln \rho(t)\bigr)$.
\par
The heat currents in the steady state $\tilde{\rho}(t)$ are time-independent and given by
\begin{equation}
\tilde{\mathcal{J}^j} = -k_B T_j\sum_{q\in \mathbf{Z}}\sum_{\{\omega\geq 0\}}\mathrm{Tr}\bigl[( \mathcal{L}^j_{q\omega}\tilde{\rho})\ln \tilde{\rho}^j_{q\omega})\bigr].
\label{curr_st}
\end{equation}
They satisfy the II-law in the form
\begin{equation}
\sum_{j=1}^M \frac{\tilde{\mathcal{J}}^j}{T_j} \leq 0
\label{IIlaw_st}
\end{equation}
while, according to the I-law
\begin{equation}
-\sum_{j=1}^M \tilde{\mathcal{J}^j} = - \tilde{\mathcal{J}} = \bar{\mathcal{P}}
\label{power_st}
\end{equation}
is the averaged power (negative when the system acts as a heat engine). Notice, that in the case of a single heat bath
the heat current is always strictly positive except for the case of no-driving when it is equal to zero.
\par
Notice, that for the constant Hamiltonian the formulas of above are also applicable after removing the index $q$, what implies also that $\sum_{j=1}^M \tilde{\mathcal{J}^j}= 0$.
\section{Appendix IV. Van Hove Phenomenon}
A natural physical stability condition which should be satisfied by any model of open quantum system is that its total Hamiltonian should be bounded from below and should possesses a ground state. In the case of systems coupled linearly to bosonic heat baths it implies the existence of the ground state for the following bosonic Hamiltonian (compare with (\ref{26}))
\begin{equation}
H_{bos} = \sum_{k}\bigl\{\omega(k)a^{\dagger}(k)a(k) + (g(k)a(k) + \bar{g}(k)a^{\dagger}(k))\bigr\}
\label{ham_bos}
\end{equation}
Introducing a formal transformation to a new set of bosonic operators
\begin{equation}
 a(k) \mapsto b(k)= a(k) + \frac{\bar{g}(k)}{\omega(k)} 
\label{trans}
\end{equation}
we can write
\begin{equation}
H_{bos} = \sum_{k}\omega(k)b^{\dagger}(k)b(k)- E_0, \ E_0 = \sum_{k}\frac{|g(k)|^2}{\omega(k)}
\label{ham_bos1}
\end{equation}
with the formal ground state $|0\rangle$ satisfying
\begin{equation}
b(k)|0\rangle =  0, \ \mathrm{for\ all}\ k . 
\label{ground}
\end{equation}
For the interesting case of infinite set of modes $\{k\}$, labeled by the $d$-dimensional wave vectors, two problems can appear:

1) the ground state energy $E_0$ can be infinite, i.e. does not satisfy
\begin{equation}
 \sum_{k}\frac{|g(k)|^2}{\omega(k)}< \infty.
\label{vanhove}
\end{equation}
2) the transformation (\ref{trans}) can be implemented by a unitary one, i.e. $b(k) = U a(k) U^{\dagger}$ if and only if
\begin{equation}
 \sum_{k}\frac{|g(k)|^2}{\omega(k)^2}< \infty.
\label{vanhove1}
\end{equation}
Non-existence of such a unitary implies non-existence of the ground state (\ref{ground}) (in the Fock space of the bosonic field) and is called \emph{van Hove phenomenon} \cite{algebricmethods}.
\par
While the divergence of the sums (\ref{vanhove}),(\ref{vanhove1})(or integrals for infinite volume case) for large $|k|$ can be avoided by putting \emph{ultra-violet cutoff} the stronger condition (\ref{vanhove1}) puts  restrictions on the form of $g(k)$ at low frequencies. Assuming, that $\omega(k)= v|k|$ and $g(k)\equiv g(\omega)$ the condition (\ref{vanhove1}) is satisfied for the following low-frequency scaling in the $d$-dimesional case
\begin{equation}
|g(\omega)|^2 \sim \omega^{\kappa} ,\ \kappa > 2- d .
\label{vanhove2}
\end{equation}
\bibliography{citeronnie,citeamikam,pub,dephc1}
\end{document}